**High efficient cyclic electron flow and functional supercomplexes in *Chlamydomonas* cells**

Pierre Joliot [a,*], Julien Sellés [a,*], Françis-André Wollman [a], André Verméglio

[a] *Laboratoire de Biologie du Chloroplaste et Perception de la Lumière Chez les Microalgues, Institut de Biologie Physico-Chimique, CNRS UMR 7141, Sorbonne Université, Paris, France*

* Corresponding authors : pjoliot @ibpc.fr (Pierre Joliot); selles@ibpc.fr (Julien Sellés)

**Abstract**

A very high rate for cyclic electron flow (CEF) around PSI (~180 $s^{-1}$ or 210 $s^{-1}$ in minimum medium or in the presence of a carbon source respectively) is measured in the presence of methyl viologen (MV) in intact cells of *Chlamydomonas reinhardtii* under anaerobic conditions. The observation of an efficient CEF in the presence of methyl viologen is in agreement with the previous results reports of Asada et al in broken chloroplasts (Plant Cell Physiol. 31(4) (1990) 557–564). From the analysis of the P700 and PC absorbance changes, we propose that a confinement between 2 PC molecules, 1 PSI and 1 cyt$b_6f$ corresponding to a functional supercomplex is responsible for these high rates of CEF. Supercomplex formation is also observed in the absence of methyl viologen, but with lower maximal CEF rate (about 100 $s^{-1}$) suggesting that this compound facilitates the mediation of electron transfer from PSI acceptors to the stromal side of cyt$b_6f$. Further analysis of CEF in mutants of *Chlamydomonas* defective in state transitions shows the requirement of a kinase-driven transition to state 2 to establish this functional supercomplex configuration. However, a movement of the LHCII antennae is not involved in this process. We discuss the possible involvement of auxiliary proteins, among which is a small cyt$b_6f$-associated polypeptide, the PETO protein, which is one of the targets of the STT7 kinase.

**Introduction**

In plants and algae, photosynthetic electron transfer operates according to two different modes. In the linear mode, electrons are transferred from water to NADP via three transmembrane protein complexes, the photosystem II (PSII), the cytochrome $b_6f$ (cyt$b_6f$) and the photosystem I (PSI). This linear electron flow (LEF) generates both an electrochemical proton gradient across the thylakoid membranes and strong reductants in the stroma via the acceptor side of PSI. In the cyclic mode, the flow of electrons between PSI and cyt$b_6f$ complexes generates an electrochemical proton gradient across the thylakoid membrane without net production of reducing equivalent. This cyclic electron flow (CEF) was first described by Arnon and coworkers on a suspension of broken chloroplasts in the presence of a catalytic amount of vitamin K [1]. Later, ferredoxin (Fd), the first soluble PSI electron acceptor, also was shown to promote CEF in broken chloroplast [2–4]. Although a non-ambiguous demonstration of this process is a difficult task mainly for technical reasons, it is largely accepted that CEF is operating *in vivo* in intact leaves or algae [5–7]. However, the nature of the components involved, the mechanistic model and the regulation of this cyclic process remain a matter of controversy.

Various types of models have been proposed for the mechanism of CEF [4, 6, 7]. In a first model, electrons are transferred from the acceptor side of PSI facing the stroma to the PQ pool via a PQ/Fd NADPH oxydo-reductase (NDH1/ NDH2). The rates of CEF associated with NDH1 in plants or with NDH2 in *Chlamydomonas*, were estimated to be equal to ~0.1 s$^{-1}$ [8, 9] or 2.5 s$^{-1}$ [10, 11] respectively. As CEF rates as high as 130 s$^{-1}$ in plants [11] or 70 s$^{-1}$ [12] in *Chlamydomonas* are reported in the literature, NDH pathways could bring only a minor contribution to CEF. In a second model, electron transfer between the acceptor side of PSI and the PQ pool would be mediated by a ferredoxin plastoquinone oxydo-reductase (FQR) [13]. However, in the absence of any genetic or biochemical support, this model remains very speculative. In the third model, stromal ferredoxin (Fd) transfers electrons from the PSI acceptor side to the cyt$b_6f$ complex according to the original Q cycle process proposed by Mitchell [14], which is distinct from the modified Q cycle [15] involved in the linear mode. The ferredoxin-NADP$^+$ oxydo-reductase (FNR) enzyme indeed was shown to bind to the stromal side of cyt$b_6f$ [16, 17]. It could provide a binding site for the attachment of Fd to the cyt$b_6f$ complex. In plants in oxidizing conditions, the reduction of the NADP pool induces a shift from linear to cyclic mode very likely associated with the binding of FNR to the cyt$b_6f$ complex [18]. In a mechanistic model [6], electron flow from Fd to site $Q_i$ was proposed to involve cyt$c_i$ [19, 20]. Finally, one should also consider the contribution of two interacting proteins, PGRL1 and PGR5, which were shown to participate in the switch between linear and cyclic modes [21, 22]. However, the maximal CEF rate being independent of the presence of PGRL1 [12], it suggests that these two proteins only contribute to the regulation of the CEF.

The structural organization of the proteins involved in CEF within the thylakoid membrane is another controversial issue. It is well established, in plant or in green algae, that PSII and PSI reactions centers are localized in different regions of the membrane, appressed and non-appressed respectively, while cyt$b_6f$ is present in these two regions. Since PQ diffusion in the appressed region is restricted to small domains (2 to 4 PSII RCs) [23, 24], LEF involves exclusively the fraction of cyt$b_6f$ complex localized in the appressed regions. On the contrary, we expect that CEF implicates predominately the fraction of the cyt$b_6f$ complex localized in the vicinity of PSI, *i.e.* in the non-appressed regions. A supramolecular organization of the electron transfer chain specific of CEF has been proposed for many years by suggesting formation of supercomplexes associating PSI and cyt$b_6f$ complexes [25, 26]. Measurements of the concentration of cyt$b_6f$ by freeze fracture analysis leads to the conclusion that about 60% of the PSI centers could be included in such supercomplexes [27]. However, single particles analysis reported candidates for such supercomplexes in *Chlamydomonas* only for a tiny minority of the isolated proteins [28]. Iwai et al [29] reported the purification of elusive supercomplexes, comprised of cyt$b_6f$ and PSI, which were able to drive CEF, in *Chlamydomonas* cells in "state 2 conditions" induced *in vivo* by the reduction of the PQ pool in presence of an uncoupler. The significance of these molecular entities was questioned subsequently since no biochemical evidence for substantial supercomplex formation between PSI and cyt$b_6f$ in a variety of experimental conditions was found [30].

Presently there is a lack of functional evidence for the presence of PSI/cyt$b_6f$ supercomplexes *in vivo*, despite the plethora of reports on the actual operation of CEF in various organisms and environmental conditions. Several of these studies concurred to suggest that the switch from linear to cyclic electron flow transition required a transition from state 1 to state 2 in *Chlamydomonas* [31-33]. Another study agreed that CEF is enhanced in reducing conditions but asserted that a transition to state 2 was not required [34]. Finally, few biochemical studies reported the existence of PSI/cyt$b_6f$ supercomplexes after placing *Chlamydomonas* cells in anaerobic conditions that induce extensive reduction of the plastoquinone pool, an experimental condition which promotes transition to state 2 [25, 29].

In the present paper, we revisit these observations in *Chlamydomonas* intact cells and demonstrate on the basis of membrane potential measurements that CEF is operating at rates as high as 210 s$^{-1}$ in anaerobic conditions in the presence of MV. This redox mediator was used in the pioneering study of Asada and coworkers [35] to promote CEF in broken chloroplasts. Using these experimental conditions, we provide evidence for a confinement between almost all PSI and part of the cyt$b_6f$ complexes that we define as functional supercomplexes. We show that the switch to this configuration upon establishment of anaerobiosis requires the activity of the STT7 kinase responsible for state transitions but does not involve lateral displacement of the LHCII antenna.

**Materials and methods**

The wild type strain of *Chlamydomonas* and several mutants, the Fud7 mutant lacking PSII [36], the *stt7-1* and *stt7-9* mutants defective in state transitions [37] and the double mutant BF4-F34 lacking PSII and most of the mobile antenna [38], were grown at 25 °C under continuous light (10 µE m$^{-2}$ s$^{-1}$) in Tris-acetate-phosphate medium (TAP). Prior to experiments, the cells were centrifuged, suspended in minimum or TAP media and then kept aerobic and dark-adapted under gentle shaking for 1 hour. We checked that the physiological state of algae, characterized by electric field changes or P700 measurements, remained stable for more than 5 hours. Owing to the variability in the kinetics of the transition to anaerobic conditions between cells batches of various cultures, averaging of experiments and comparison of light-induced kinetics of membrane potential, P700 and PC were performed on a same batch of algae.

In most experiments, oxidized MV was introduced in large excess (6 mM) to ensure that a large amount remains oxidized even after a long anaerobic incubation.

Absorption changes were measured with a JTS spectrophotometer (Biologic). Absorption changes were measured in a cuvette (8 mm diameter and 2.5 mm thick) placed horizontally, therefore perpendicular to the vertical analyzing and actinic beams. This arrangement allows sedimentation of algae without significant drift in absorption changes. For measurements in anaerobic conditions, we used a closed cuvette in which the respiration-mediated anaerobic conditions are spontaneously reached after 3 to 15 min incubation depending upon the physiologic state and the concentration of algae. For aerobic conditions, a similar cuvette was used but its bottom was an O$_2$-permeable Teflon sheet, 0.02 mm thick. Actinic lights were provided by an orange LED pulse of 160 ms peaking at 630 nm and by a saturating laser flash of <10 ns duration (690 nm) synchronized to the onset of the orange pulse.

Cells were placed at time zero in a closed cuvette and then submitted to this illumination given every 1 min. Light is interrupted for 200 µs after 150 ms illumination and the absorption is sampled after 100 µs of dark. The wavelength of the detecting beam is alternatively 520 nm and 546 nm. The membrane potential change is computed as the difference ΔI/I 520 (t) – ½ (ΔI/I 546 (t-1) + (ΔI/I 546 (t+1)/2) where t is the time of incubation in minutes.

The PSI photochemical rate constant, k$_{iPS1}$, under continuous illumination corresponds to the initial rate of increase in transmembrane potential. It is worth pointing out that the PSI rate constant depends on the size of the PSI antenna and on the algal concentration, owing to the absorption of the 630 nm actinic light within the cuvette. P700$^+$ and PC$^+$ reduction kinetics were measured during the dark period that follows the end of the 160 ms light pulses. P700 redox changes were computed as the absorption difference ΔI/I 700 nm - ΔI/I 730 nm. PC redox changes were measured by the absorption changes ΔI/I 573 nm or ΔI/I 730 nm, the two isobestic points of P700 redox absorption changes. The base line was measured after more than 10 s after the end of the pulse, *i.e.* after completion of the reduction of P700$^+$ and PC$^+$.

## Results

*Cyclic electron flow measured in the presence of methyl viologen*

The approach most generally used to quantify the rate of CEF is the measurement of the membrane potential changes at the end of a period of illumination in the absence of PSII activity. For *Chlamydomonas* cells, this can be achieved either by PSII inhibition by a preillumination of the wild type strain in the presence of hydroxylamine (HA) and DCMU, or by using a mutant devoid of PSII. In the following, we chose the latter option with the Fud7 mutant, fully lacking PSII cores due to the deletion of the *psbA* gene coding for the D1 subunit [36], but similar results were obtained when using the wild type strain after PSII inhibition.

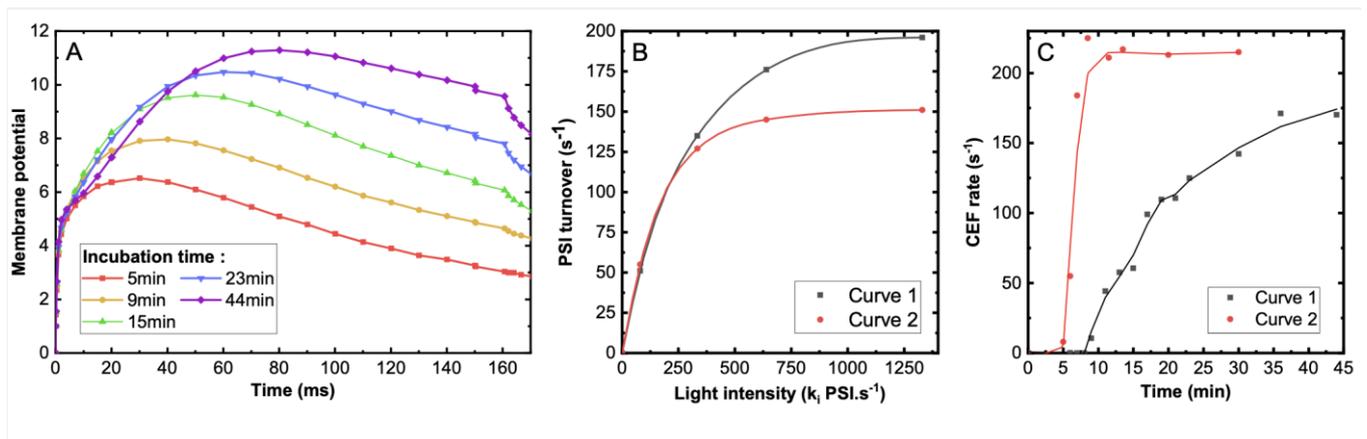

**Fig. 1**: *Membrane potential changes and CEF rates induced by saturating light pulses on a Chlamydomonas Fud7 mutant. Cells were placed at time zero in a closed cuvette and then submitted to a saturating continuous illumination ($k_{iPSI}$ ~4900 $s^{-1}$) of 160 ms light pulse every 1 min.*
*A: Kinetics of light-induced membrane potential by a series of saturating light pulses in minimum medium after 5, 9, 15, 23, and 44 min of incubation.*
*B: Initial rates of the membrane potential decay measured after a 160 ms pulse of various intensities. Curve 1: Initial rates computed in a time interval 100 µs - 6 ms. Curve 2: Initial rates computed in time interval 1.5 ms - 6 ms.*
*C: Rate of cyclic flow measured at the end of each pulse as a function of the time of incubation.*
*Cells suspended in a closed cuvette in minimum medium (curve 1) or in TAP medium (curve 2). CEF rate is the initial slope computed after extrapolation of the decay kinetics computed in a time interval 1.5 ms - 10 ms according to a second order polynomial fit.*

To study the transition of photosynthetic electron flow from the LEF to CEF mode, the Fud7 mutant was incubated in a closed cuvette and subjected to a series of 160 ms saturating light pulses in the presence of 6 mM oxidized MV. A saturating laser flash that induces a single PSI turnover is synchronized at the onset of the illumination. After 150 ms illumination, light is interrupted during 200 µs and the absorption sampled after 100 µs of dark and membrane potential decay is analyzed at the end of the 160 ms pulse. Fig. 1A shows the kinetics of the transmembrane potential changes induced by 160 ms light pulses measured on algae suspended in minimum medium. Algae were incubated in a closed cuvette for 44 min during which they switch from an oxygenic to an anaerobic environment. The large increase in membrane potential during illumination is linked exclusively to PSI and cyt $b_6f$ activities. At the end of the pulse, the rate of membrane potential decay is the sum of the rate of proton leaks through the membrane via the ATPase and the rate of back reactions within PSI. A more detailed analysis of the kinetics of the membrane potential changes measured 10 ms before and 20 ms after the end of the pulse is shown in supplemental material (Fig. A supp. mat.). During a short period of incubation (<8 min), the algae stay in oxygenic conditions and the decay kinetics of the membrane potential does not change at the end of the light pulse. This means that the PSI turnover or back reactions were close to zero, *i.e.* no measurable CEF, as already reported in [12]. For incubation times longer than 8 min, we observe an increase in the rate of decay of the transmembrane potential after the end of the pulse, which indicates a significant PSI turnover due a combination of CEF and back reactions within PSI. In the range of light intensity used in fig. 1, the kinetics of back reaction completed in less than 1 ms, the amplitude of which is an increasing function of the light intensity [11]. The more likely electron acceptors involved in this process are the reduced forms of the Fx and of the two quinones acceptors $A1_A$ and $A1_B$.

To better discriminate between the contributions of CEF and back reactions we measured the turnover on algae suspended in minimum medium as a function of the light intensity (Fig. 1B). As discussed in [11, one expects that back-reactions are an increasing function of the light intensity while CEF should reach a maximum level limited by the maximum cyt$b_6f$ turnover. Experiments were performed after more than 35 min incubation, *i.e.* when a maximum PSI turnover is reached. In a first deconvolution, the initial rate of the membrane potential decay kinetics is determined using a second order polynomial fit in the time interval 100 μs to 6 ms (Fig. 1B, curve 1). The observation that curve 1 is an increasing function of light intensity reflects a contribution of back-reactions that occur in the 100 μs to 1 ms range [10]. For a second deconvolution, we used a time interval between 1.5 ms to 10 ms (Fig. 1B, curve 2). The plateau level observed for the rates in curve 2 for intensities larger than 600 $s^{-1}$ implies that they could be ascribed to CEF exclusively. We therefore used this second deconvolution procedure to eliminate the contribution of the back-reactions to the decay kinetics (Fig. A supp. mat.). The rate of CEF was then estimated by measuring the difference between the rate of membrane potential changes during the last 10 ms of illumination and the rate of the initial decay measured on the extrapolated kinetics. As shown on fig. 1C, after a lag phase during the first 8 min of incubation in the cuvette, the rate of CEF progressively increases and reaches a maximum value of ~180 $s^{-1}$ after ~35 min of incubation. When cells are maintained in aerobic conditions (see material and methods), the rate of CEF remains close to zero even for times of incubation longer than 50 min (data not shown). Thus, the shift from linear to cyclic mode is triggered by anaerobic conditions.

In contrast to the previous experiments above, when Fud7 cells were resuspended in TAP medium, a maximum CEF rate of ~210 $s^{-1}$ was already reached in less than 10 min incubation (Fig.1C) (deduced from experiments shown in Fig. B, supp. mat.). We attribute these fast rates and fast transitions to the more reducing conditions imposed by the incubation in TAP medium. The differences between the results measured in minimum and TAP media illustrate the dependence of both the CEF rates and the time of transition from LEF to CEF on the physiological state of the algae and more specifically on the redox poise within the stromal compartment.

In Fig. C supp. mat., we analyzed the kinetics of membrane potential decay over a period of 45 ms, following a 150 ms pulse of saturating light on a Fud7 mutant suspended in TAP medium (blue curve). This kinetics display a first phase completed in about 10 ms, as already observed in plants leaves at the end of pulses of saturating light [40, 41]. This phase was interpreted assuming that, above a critical value of the electrochemical proton gradient, the rate of ATP synthesis at the level of the F1 complex has reached its maximum value but the rotation rate of the electrical rotor F0 still increases [40, 41]. This process induces a proton leak not coupled to ATP synthesis (proton slip). The 10 ms decay phase is followed by a progressive decrease in the decay rate completed in a few tenths of ms.. This decrease reflects a slowdown of the proton pumping process associated with the reduction of the PSI donors at site $Q_o$ of the cyt$b_6f$ complex. The red curve (Fig. C supp. mat.) is computed by extrapolation to time 0 of the kinetics measured in the 10 to 45 ms time range. This initial rate gives a rough estimate of the rate of the CEF coupled to ATP synthesis (about 50 s$^{-1}$), much less than the maximum CEF rate measured after extrapolation of the decay phase in the 1.5 to 10 ms time interval (210 s$^{-1}$) (Fig. 1C).

To check if the large CEF rate measured in the presence of MV is associated with cyt$b_6f$ turnover, we analyzed the effect of DNP-INT, an inhibitor of site $Q_o$ of this complex [42], on this rate for a WT strain of *Chlamydomonas* (Fig. D supp. mat.). In the absence of inhibitor, PSI turnover measured is 210 s$^{-1}$, *i.e.* similar to that measured in on the Fud7 mutant. Addition of DNP-INT (25 µM) inhibits PSI turnover by a factor of 50 showing that cyt$b_6f$ is actually involved in the CEF measured in the presence of MV. Moreover, the large slowdown of the rate of membrane potential decay implies that this compound also inhibits the ATPase.

*Analysis of kinetics of membrane potential increase at the onset of the illumination*

Figure 2 shows the membrane potential increase at the onset of the illumination during evolution from an oxygenic to an anaerobic environment. Under oxidizing conditions (5 min incubation), electron transfer reactions between cyt$b_6f$ and PSI occur exclusively according to the linear mode and the membrane potential increase is associated only with the oxidation of primary and secondary PSI donors. Under these conditions, the kinetics of the increase in membrane potential display a fast phase, completed in less than 4 ms followed by a slower phase in the range of tens of ms (Fig. 2A, curve 1). After correction for the slow decay of the membrane potential associated with proton leak through the membrane (Fig. 2A, curve 2), the maximal value of the membrane potential increase corresponds to the oxidation of 7.8 electron donors per PSI centers. These donors included P700, PC, cyt$f$, Rieske protein and a small fraction of reduced PQ rapidly accessible. This fraction of reduced PQ is also responsible for the slow phase of the membrane potential increase induced by a short saturating flash and is equal to ~0.5 electron equivalent (Fig. 2A, insert). On this basis, we estimate to 7.8 – 0.5 = 7.3 electron equivalents the total number of P700, PC, cyt$f$, and Rieske protein per PSI, present in the thylakoids. Oxidation of about half of these donors (~4 electron equivalents) is completed in ~3 ms while the second fraction (~3.3 electron equivalents) requires up to 50 ms (Fig. 2A, curve 1). The fast oxidation of the secondary donors (P700, PC, cyt$f$ and Rieske protein) requires their close vicinity with the PSI centers. We therefore propose that these secondary donors are localized in the non-appressed region. On the other hand, the slow phase corresponds to the oxidation of PC, cyt$f$ and Rieske protein in the appressed region which is limited by the rate of diffusion of PC between the appressed and the non-appressed regions. Between 5 and 8 min of incubation, most of the PSI centers still operate according to the linear mode as determined by the rate of the membrane potential decrease at the end of the illumination (Fig. A, supp. mat.).

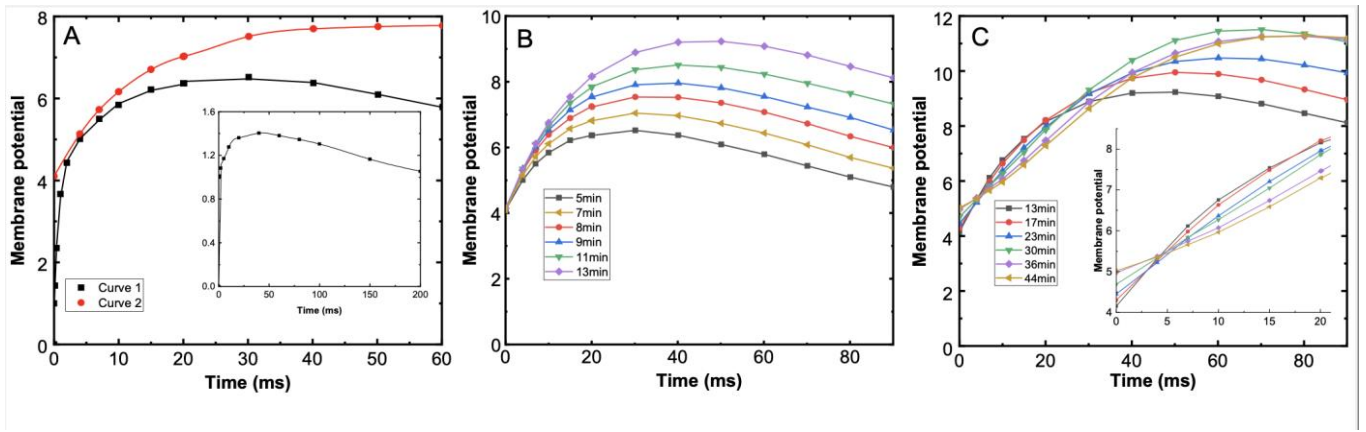

**Fig. 2:** *Kinetics of membrane potential decay measured after a 160 ms pulse on a Chlamydomonas fud7 mutant. Cells were suspended in minimum medium in the presence of MV (6 mM). The light intensity corresponds to $k_{iPSI}$ = 3000 $s^{-1}$.*
*A: Time of incubation: 5 min. Curve 1: kinetics plotted in the time interval 0 to 60 ms. Curve 2: kinetics of the slow phase after subtraction of the slow linear decay in the 50 ms to 80 ms interval extrapolated to time zero for time shorter than 5 ms. Insert: kinetics of membrane potential changes induced by a short saturating flash after 5 min incubation.*
*B: Kinetics of the slow phase for cells incubated from 5 min to 13 min. The slow phases are plotted between 4 ms and 90 ms and extrapolated to time zero.*
*C: Same as part B but for cells incubated from 13 min to 44 minutes. Insert: emphasis of the kinetics in the first 20 ms.*

During this incubation period, due to a partial reduction of the PQ pool the amplitude of the slow phase progressively increases (Fig. 2B). For longer times of incubation, a progressive shift from the linear to the cyclic mode occurs (see Fig. 1C) and the amplitude of the fast phase progressively increases from 4 to 5 electron equivalents showing that one additional PSI secondary donor rapidly transfers an electron to P700 (Fig. 2C). Increase of the fast phase correlates with a decrease of the initial slope of the slow phase and, after 36 min of incubation, the slow phase displays a lag phase completed in 10 ms (Fig. 2C, insert) that we ascribe to a transitory inhibition of the cyt$b_6f$ complex. This transitory inhibition is discussed in the supplementary material section (Fig. E, supp. mat.).

As shown on fig. D supp. mat., about 5 PSI donors are rapidly oxidized in the presence of DNP-INT showing that this inhibitor does not perturb electron transfer between secondary donors and P700. The fast oxidation of 5 PSI electron donors reflects a proximity between these electron carriers. The electron transfer rates and the thermodynamic equilibration between P700 and PC are studied in the following paragraph.

*Analysis of P700 and PC redox changes at the end of the illumination period*

P700 redox changes were performed in a closed cuvette according to a protocol similar to that used in fig. 1. Light pulses of 160 ms duration are given 1 min apart on cells suspended in minimum medium for times of incubation ranging from 5 min to 35 min. In oxygenic conditions (less than 8 min incubation), the pulse induces the full oxidation of P700 (Fig. 3A) as already reported in [12]. Following the light pulse, P700$^+$ is reduced according to a slow monophasic process (t$_½$~620 ms) (Fig. 3A). We assume that this slow process is associated with electron transfer from reduced PSI acceptors via NDH2 and then to P700 via the Q$_o$ site of the cyt$b_6f$ complex. For longer times of incubation, an increasing fraction of P700 remains reduced at the end of the light pulse. The residual oxidized fraction is reduced in two distinct phases: a fast phase completed in ~30 ms with a roughly constant halftime of ~4 ms (Fig. 3B) and a second phase with halftimes decreasing, from 600 ms to 200 ms, when the incubation time increases (Fig. 3A). Thus, two routes for reduction of oxidized P700 should be considered in these experimental conditions. Note that at the end of the saturating pulse for long times of incubation only a small fraction of reduced P700 (~30%) are in an active form due to the large ratio between the photochemical constant ($k_{PSI}$ = ~4700 s$^{-1}$) and the CEF rate (~160 s$^{-1}$). Thus, most of the centers with a reduced P700 include PSI acceptors (from A1 to $F_A F_B$) in reduced states that prevent charge stabilization. The pool of reductants stored in PSI acceptors (~5 electron equivalents) provides the electrons transferred via the Q cycle to the oxidized PSI donors.

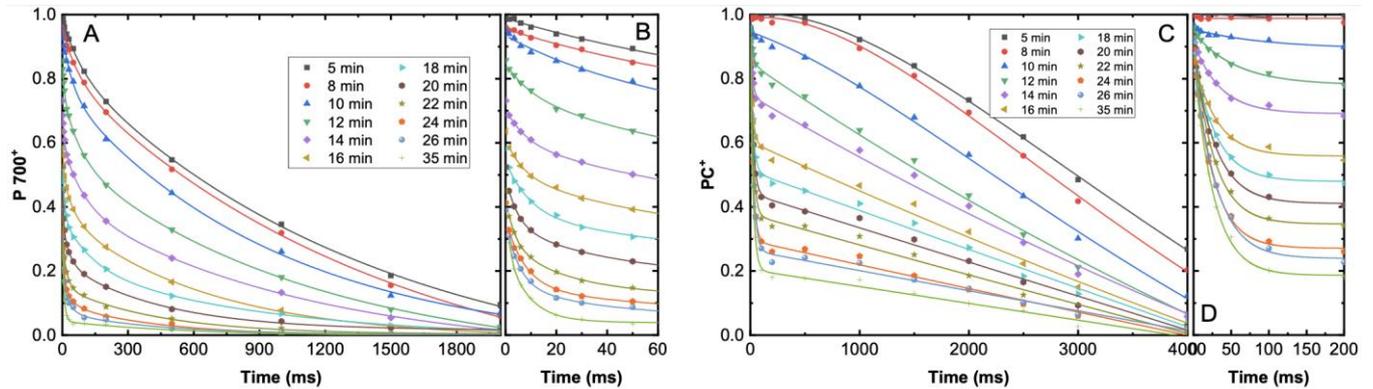

**Fig. 3:** *Kinetics of P700$^+$ and PC$^+$ reductions measured at the end of a 160 ms pulse of saturating light as a function of the time of incubation.*

*Fud7 mutant cells were suspended in minimum medium in the presence of MV 6 mM. Intensity of the light pulse: $k_{iPSI}$=1600 s$^{-1}$.*
*A: Kinetics of P700$^+$ reduction measured between 0 and 2000 ms.*
*B: Same as part A but between 0 and 60 ms.*
*C: Kinetics of PC$^+$ reduction measured between 0 and 5000 ms.*
*D: Same as part C but between 0 and 200 ms.*

Two routes are also observed for the kinetics of the reduction of the oxidized PC. In oxygenic conditions, reduction of PC$^+$ is a slow process completed in ~5 s (Fig. 3C). The kinetics displays a lag phase (~500 ms duration) explained by the lower value of the mid-point redox potential of PC compared to that of P700. For longer times of incubation, the kinetics of PC$^+$ reduction display two distinct phases: a fast phase whose amplitude is an increasing function of the time of incubation while its halftime is roughly constant (~18 ms) (Fig. 3D) and a slow phase, the amplitude of which decreases with the time of incubation while its halftimes shorten from 3 s to 1.4 s during the course of incubation (Fig. 3C). Similar kinetics for PC$^+$ reduction is measured at 573 nm or 730 nm, two isobestic points for P700 redox absorption changes (Fig. F, supp. mat.).

The biphasic kinetics observed for both P700$^+$ and PC$^+$ reduction suggests that these carriers are included in two distinct compartments operating according two different modes of electron transfer. These two compartments are clearly illustrated when plotting the concentration of PC$^+$ versus that of P700$^+$ during the reduction phase that follows the illumination period (Fig. 4A).

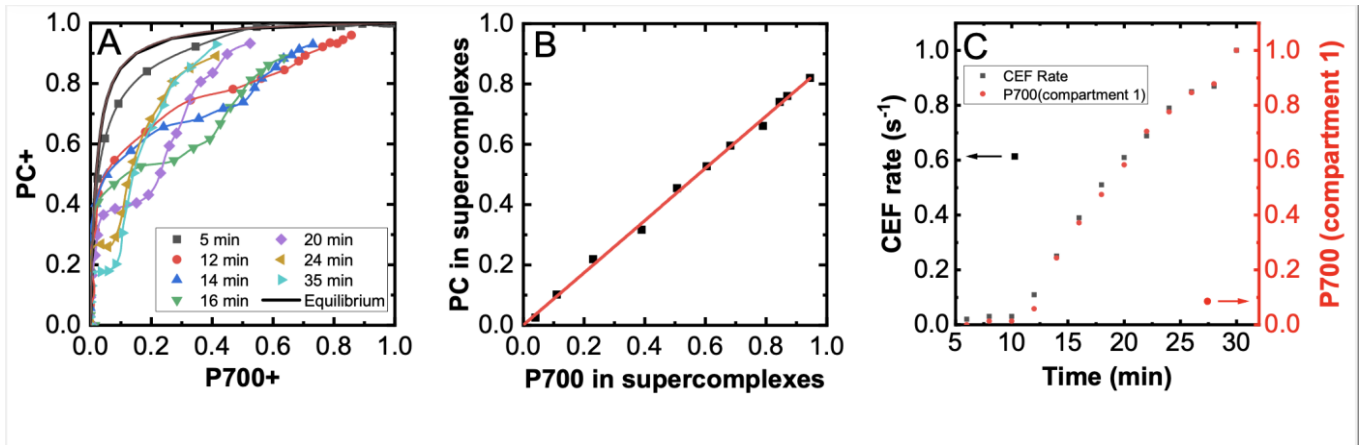

**Fig. 4**: *A: Function PC$^+$ = f(P700$^+$) in function of the incubation time computed from the data of Fig. 3.*
*Dark line: theoretical function PC$^+$ = f(P700$^+$) computed for full thermodynamic equilibration between P and PC.*
*B: Fraction of PC and P700 included in compartments 1 for various times of incubation.*
*The fraction of P700 included in compartment 1 (functional supercomplexes) is the difference between the total amount of P700 and the fraction of P700 included in compartment 2. The total amount of P700 corresponds to the maximum P700 measured in oxygenic conditions while the amount of P700 included in compartment 2 is determined by extrapolation to time zero of the slow phase of P700 reduction (Fig. G, supp. mat.).*
*C: Rate of cyclic flow and fraction of P700 included in compartment 1 as a function of the time of incubation. Black curve: CEF rates. The maximum CEF rate reached after 35 min incubation (~175 s$^{-1}$) is normalized to 1. Red curve: fraction of P700 included in compartment 1.*

In oxygenic conditions (5 min incubation), the relation PC$^+$ = f(P700$^+$) is monotonous and close to thermodynamic equilibration (Fig. 4A). The small deviation from full thermodynamic equilibration is due to the slow equilibration between the PC and the cyt$b_6f$ complexes localized in the non-appressed region and in the appressed region (see preceding paragraph). For longer times of incubation (> 12min), two distinct parts are observed in the relation PC$^+$ = f (P700$^+$) showing that electron carriers are included in two compartments far to be in thermodynamic equilibrium. On the right part of the function, the two carriers are included in a compartment where they are rapidly reduced via CEF. We propose that, in this compartment (denoted 1), PC molecules are trapped in mobility-restricted domains, here defined as "functional supercomplexes", including PSI and cyt$b_6f$, thus preventing their thermodynamic equilibration with those freely diffusing in the lumen. In the second compartment (denoted 2), (left part of the function PC$^+$ = f(P700$^+$), the slow reduction of PC$^+$ and P700$^+$ is due to electrons which are transferred via NDH2 (linear process) to the PQ pool to cyt$b_6f$ and PSI which are not part of functional supercomplexes. In this compartment, PC diffuses freely in the lumen. The rate constant of this slow reduction process progressively increases with the time of anaerobic incubation, very likely due to the increase in reduced substrates for the PQ reductase (Fd and NaDPH). The total amount of P700 corresponds to the maximum P700 measured in oxygenic conditions while the amount of P700 included in compartment 2 is determined by extrapolation to time zero of the slow phase of P700 reduction (Fig. G, supp. mat.). Therefore, the fraction of P700 included in compartment 1 (functional supercomplexes) is the difference between the total amount of P700 and the fraction of P700 included in compartment 2. In Fig. 4B, we have plotted [P700$^+$] against [PC$^+$] for the fraction of donors included in functional supercomplexes (compartment 1). Concentrations of PC$^+$ and P700$^+$ are linearly related showing that the stoichiometric ratio PC / P700 stays constant during the incubation period. Since the fraction of P700 included in functional supercomplexes is proportional to the CEF rate (Fig. 4C), CEF occurs only in these functional supercomplexes. The stoichiometry between primary and secondary donors P700, PC, cyt*f* and Rieske protein and the structural organization of the supercomplexes stay constant whatever their concentration. After more than 35 min incubation, the maximum fraction of P700 trapped in supercomplexes is about 95 % (see Fig. 4B). The fraction of PC included in supercomplexes is ~80% while the remaining freely diffuses in the lumen. This fraction varies from batch to batch. It is more than 95% in the experiments of fig. F (supp. mat.). Therefore, most of the PC molecules, including a large part of those previously localized in the appressed regions are trapped between cyt$b_6f$ and PSI after a long period of incubation. The trapping of one additional PC molecule in the supercomplexes is associated with the increase from 4 to 5 in the number of electrons rapidly transferred to P700 during the transition from linear to cyclic flow (Fig. 2C). Taking into account that ~5 donors are rapidly oxidized, a reasonable assumption is that each functional supercomplex contains 1 PSI, 2 PC molecules, 1 cyt*f* and 1 Rieske protein.

*Dependence of supercomplex formation upon state transitions*

To determine the role of state transitions on the formation of functional supercomplexes, CEF rates were measured on suspension of the mutant *stt7-1* [37], defective in state transitions, during the transition from aerobic to anaerobic conditions (Fig. H, supp. mat.). Algae were submitted to a series of saturating light pulses of 160 ms duration, in minimum medium in the presence of MV (6 mM). PSII was inactivated by addition of DCMU (30 μM) and hydroxylamine (2 mM). In oxygenic conditions - less than 8 min incubation in the measuring cuvette - the rate of CEF was close to zero in the Fud7 mutant. Rates of CEF displayed an increase in a time range similar to that measured on the Fud7 mutant (see Fig. 1C) up to a maximum value of 117 s$^{-1}$. Using a similar approach, we measured a CEF rate of 78 s$^{-1}$ in another allelic mutant, the *stt7-9* mutant. These levels remained significantly lower (30% and 43% respectively) than that measured in the Fud7 mutant (180 s$^{-1}$) or in the WT strain in the presence of DCMU and HA (177 s$^{-1}$) in state 2 conditions (Deduced from Fig. D supp. mat.).

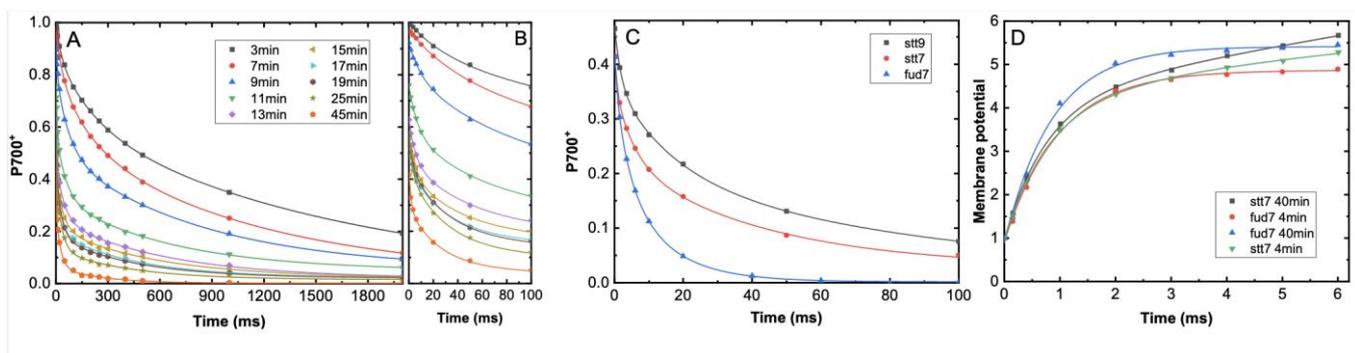

**Fig. 5:** *Kinetics of P700$^+$ reduction and membrane potential increase measured on the stt7-1, stt7-9 and Fud7 mutants.*

A: *Kinetics of P700$^+$ reduction measured for various times of incubation between 0 and 2000 ms.*

B: *Same as part A but between 0 and 60 ms.*

C: *Comparison of the kinetics of P700$^+$ reduction measured on the stt7-1, stt7-9 and Fud7 mutants. Cells incubated for 35 min in anaerobic conditions.*

D: *Comparison of the kinetics of membrane potential increase measured at the onset of illumination for the Fud7 and the stt7-1 mutants. Cells were incubated for 4 min or 40 min in anaerobic conditions.*

Fig. 5A shows the kinetics of P700$^+$ reduction in the *stt7-1* mutant at the end of the 160 ms light pulse, as a function of the time of incubation. After 3 min incubation, P700$^+$ reduction kinetics is a slow process close to that measured for the Fud7 mutant in aerobic conditions. For increasing times of incubation, the fraction of reduced P700$^+$ at the end on the pulse and the rates of P700$^+$ reduction progressively increase and reach steady state levels after ~45 min of incubation. Kinetics of P700$^+$ reduction is a multiphasic process and does not display the fast phase completed in 30 ms (Fig. 3B) that we observed in the Fud7 mutant and tentatively ascribe to the formation of functional supercomplexes. In figure 5C, we compared the kinetics of P700$^+$ reduction measured in the *stt7-1*, *stt7-9* and Fud7 mutants after 35 min incubation. At the end of the pulse, a similar amount of P700$^+$ is observed for the three strains (~40%) but the kinetics largely differs. While the kinetics measured in the Fud7 mutant is nearly exponential, the kinetics measured on the *stt7-1* and *stt7-9* mutants are slower and reflect a multiphasic process. These rates of P700$^+$ reduction qualitatively correlate with the CEF rates determined by the membrane potential decay as described above on the same batch of algae (180 s$^{-1}$, 110 s$^{-1}$ and 78 s$^{-1}$ for the Fud7, *stt7-1* and *stt7-9* mutants respectively).

On the basis of membrane potential measurements, we thus conclude that anaerobic incubation of the *stt7-1* and *stt7-9* mutants, although promoting CEF, does not induce the formation of these functional supercomplexes. In order to determine if the migration of the PSII antenna induced by state transition is required for the observation of a functional supercomplex configuration, we analyzed the kinetics of P700$^+$ reduction in a double mutant BF4-F34 [38] that lacks PSII centers and most of the mobile antenna. Kinetics of P700$^+$ reduction is close to those observed for the Fud7 mutant and clearly completed after 30 ms of dark (Fig. I, supp. mat.). Our overall conclusion is that the switch to a supercomplex functional configuration requires state 2 conditions and an active STT7 kinase but not the movement of the mobile LHCII antennae.

As the capability of PSI to oxidize rapidly 5 electron donors is the consequence of the formation of a functional supercomplex (see &3.2), we used this property to bring further support to the requirement of state 2 for the switch to this supercomplex functional configuration. In a first set of experiments, we compared the kinetics of membrane potential increase at the onset of illumination for the *stt7-1* and the Fud7 mutants (Fig. 5D). In oxidizing conditions, the fast phase was of similar amplitude in the two mutants (~4 electron equivalents). For the *stt-7-1* mutant, no significant change in the amplitude of the fast phase occurred after transition from aerobic to anaerobic conditions (4 electrons transferred), while for the Fud7 mutant, the number of electrons rapidly transferred increased from ~ 4 to ~5 (Fig. 5D). The fact that, only 4 electron equivalents were rapidly photo-oxidized in the stt7-1 mutant even under anaerobic conditions implies the absence of functional supercomplexes formation in agreement with our previous conclusion (§ 3.2). In addition, kinetics measured in the *stt7*-1 mutant, placed under anaerobic conditions, did not display the lag phase observed with the Fud7 mutant showing that the transitory inhibition of the cyt$b_6f$ complex is only observed upon formation of these functional supercomplexes (Fig. E part 2, supp. mat.).

*Analysis of CEF in the absence of MV*

To better understand the effect of MV on CEF, we compared CEF rates in the presence or absence of this mediator (Fig. 6A). In minimum medium and in the presence of MV, a maximum CEF rate of 160 s$^{-1}$ was reached in ~16 min (black curve), a time shorter than reported for the experiment of figure 1C, very likely due to a larger reducing power within the cells (see Discussion). In the absence of MV (red curve), the maximal CEF rate was 100 s$^{-1}$ under anaerobic conditions. In another set of experiments, the maximum CEF rates measured in the presence or absence of MV were ~180 s$^{-1}$ and ~130 s$^{-1}$ respectively (Fig. J, supp. mat.). CEF rates in the absence of MV are therefore between 40 and 30% lower than those measured in the presence of MV. During a 5 s illumination, these rates decrease slightly faster (~180 s$^{-1}$ to ~118 s$^{-1}$) in the presence of MV than in the absence of MV (~130 s$^{-1}$ to ~98 s$^{-1}$) (see Fig. J, supp. mat.).

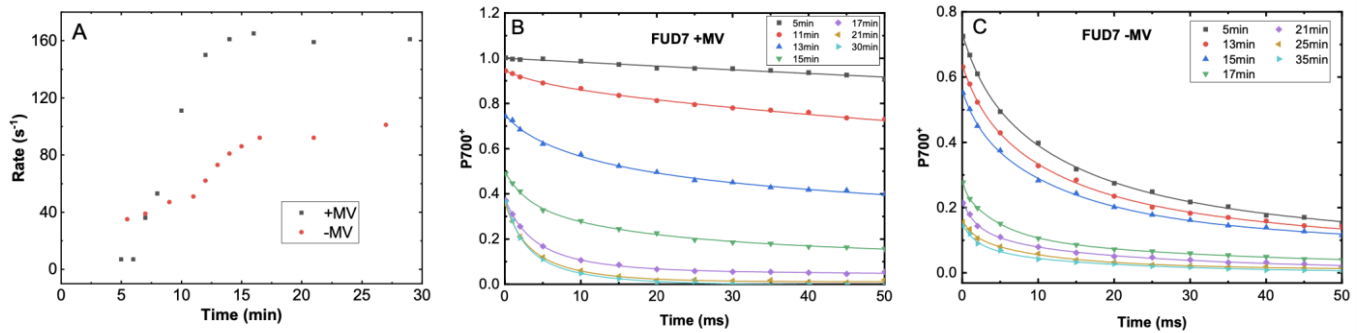

**Fig. 6:** *Comparison of the CEF rates and kinetics of P700$^+$ reduction in the presence of the absence of MV. Fud7 cells suspended in minimum medium.*
*A: CEF rates measured at the end of each pulse as a function of the time of incubation for a suspension of fud7 mutant in the presence (black curve) or the absence (red curve) of MV (6 mM).*
*B: Kinetics of P700$^+$ reduction measured at the end of a 160 ms pulse of saturating light as function of the time of incubation in the presence of MV (6 mM).*
*C: Same as part B but in the absence of MV.*

We wondered whether these differences in CEF rates were related to the absence of functional supercomplexes in absence of MV.

We addressed this question by measuring the rates of reduction of P700$^+$. In fig. 6 we compared the kinetics of P700$^+$ reduction following a 160 ms pulse of saturating light (same protocol as in fig. 3) measured in the presence (Part B) or absence (Part C) of MV for long times of incubation. The kinetics of P700$^+$ reduction displayed a fast phase completed in ~30 ms both in the presence or the absence of MV, suggesting that formation of functional supercomplexes also occurs in the absence of this mediator. The presence of a slower phase of small amplitude in the absence of MV, which may explain the lower CEF we measured, is discussed below. After a 35 min incubation, the fraction of P700$^+$ at the end of the pulse was about 2.5 times lower in the absence than in the presence of MV (0.15 versus 0.37 respectively). The larger fraction of reduced P700 in the absence of MV suggests that the rate of electron transfer from PSI acceptors to the stromal side of cyt$b_6f$ within the supercomplex is a slower process when mediated by Fd alone.

**Discussion**

In a recent study [12], a maximum rate of CEF of ~70 s$^{-1}$ was measured at the onset of illumination in the presence of DCMU on several *Chlamydomonas* strains. In the present study, addition of MV known as an efficient electron acceptor at the level of the stromal side of PSI, prevented the occurrence of CEF in aerobic conditions as expected. However, in anaerobic conditions, we measure a very efficient CEF (up to 210 s$^{-1}$) in the presence of MV in agreement with the earlier report by Asada et al [35] who described the occurrence of a MV-dependent cyclic electron flow on broken or intact spinach chloroplast in anaerobic conditions. Based on the kinetics of reduction of P700$^+$, they documented that reduced MV is the reductant of P700$^+$ induced by a saturating light pulse. In addition, this P700$^+$ reduction was inhibited by 2,5-dibromo-3-methyl-6-isopropyl-p-benzoquinone, showing that the MV-mediated cyclic electron flow involved the cyt$b_6f$ complex. The less efficient CEF process and the higher reduction level of P700 we observed in intact cells of *Chlamydomonas* in the absence of MV suggest that electron transfer from the donor side of PSI to cyt$b_6f$ is a slower process when mediated by Fd than by MV. In addition, in the absence of MV, a slower phase that follows the 30 ms phase suggests that a small fraction of reduced Fd could escape from the supercomplex and be released on the stromal side of the membrane. In any case, one has to remember that the very high rates of CEF measured in this article are not coupled to ATP synthesis due to the proton slip at the level of the ATPase (Fig. C supp. mat.). This may prevent the deleterious effect of an excess of light, mediated by a sustained acidification of the lumen because ATP synthesis becomes a limiting step for photosynthesis at saturating light intensities.

Transitions from the linear to the cyclic mode occur in the same time range in the Fud7 and *sst7*-1 mutants or in Fud7 in presence or absence of MV when placed in minimum medium (~30 min). These results suggest that a common process controls the shift from linear to cyclic mode in the presence or absence of kinase. In plants in state I conditions, it was proposed [18] that the shift from the linear to cyclic mode was associated with the binding of FNR to the stromal side of the cyt$b_6f$ complex. This process is triggered by the reduction of the NADP pool. We propose that, in *Chlamydomonas* as well, CEF is also triggered by the reduction the NADP pool and the binding of FNR to the cyt$b_6f$, in the presence or absence of MV. This process is a prerequisite for the formation of supercomplexes occurring only in the presence of the STT7 kinase. As the rate of reduction of NADP depends on the reducing power within the cells, this would explain why a much faster transition is observed when algae are suspended in TAP medium.

The high rates of CEF, the very fast electron transfer measured between PC and P700$^+$ and their thermodynamic equilibration imply a proximity between PC, PSI reaction centers and cyt$b_6f$ complexes in anaerobic conditions. MV may facilitate the mediation of electron transfer from PSI acceptors to the stromal side of cyt$b_6f$ within this confinement but is not necessary for the building up of this supramolecular confinement. Moreover, this confinement is not mandatory for an efficient CEF as shown by the behavior of the mutants lacking an active kinase.

The present experiments show that the vast majority of PSI reaction centers can be confined with cyt$b_6f$ complexes. They associate in a fixed stoichiometry of 1 PSI, 1 cyt$b_6f$ and 2 PC. Given that cyt$b_6f$ complexes are present as dimers in the thylakoid membranes, we are led to consider that 1 cyt$b_6f$ dimer interacts with 2 PSI and 4 PC in what we defined as a functional supercomplex. Such a functional association implies the trapping of the 4 PC within domains that group 1 cyt$b_6f$ dimer and 2 PSI. These functional entities could correspond to those supercomplexes that were proposed, based on biochemical studies providing evidence for copurification of a fraction of PSI with cyt$b_6f$ complexes [25-29]. Beyond the controversy regarding the actual significance of these biochemical preparations [30], we note that functional supercomplexes may as well result from more labile interactions due to a confinement between PSI, cyt$b_6f$ and PC in microdomains produced by a particular lipid organization or by a bridging protein trapping PC between PSI and cyt$b_6f$ at the luminal surface of the thylakoid membranes. On the basis of single particle analysis and single molecule fluorescence measurements, Steinbeck et al [28] proposed a structural model in which 1 cyt$f$ bound to PSI complex would be localized in the vicinity of PSAF. In this model, 2 PC may be trapped in the small cavity separating the 2 PC binding sites of cyt*f* and PSI. Only a small movement of the PC molecules trapped in this cavity would be required for electron transfer from cyt*f* to P700$^+$.

The fact that such functional supercomplexes may not have a supramolecular counterpart amenable to biochemical purification is not unprecedented. For example, *Rhodobacter sphaeroides* also houses functional supercomplexes between 2 bacterial reaction centers, 1 cyt*bc₁* complex and 1 cytochrome $c_2$ [43] but no biochemical supramolecular counterparts have yet been identified. Conversely, the supramolecular organization of PSII reaction centers in dimers, which is well documented by biochemical and crystallographic studies [44, 45] does not correspond to any known functional entity endowed with properties that would not be present in each monomer independently.

State transitions were originally recognized as a process governing antenna redistribution between the two photosystems to cope with their unbalanced excitation [46]. It was later shown that it is triggered by a reversible phosphorylation of the peripheral antenna in plants and green algae, under the control of a kinase active only upon reduction of the PQ pool. It was further demonstrated that the physiological significance of state transitions extended far beyond an antenna redistribution since transition to state 2 also was promoted by a drop in intracellular ATP [47] accompanied by an increased concentration of cyt$b_6f$ in the PSI-enriched membrane domains [31]. These characteristics of state 2 were taken as indicative of a possible function of this state in the establishment of a CEF configuration aimed at restoring higher ATP levels [48]. Still the role of state transitions in the regulation of CEF has remained highly controversial. Wollman and Bulté [25], using parabenzoquinone treatment to stabilize state I or state 2, originally observed that a large fraction of cyt$b_6f$ complexes was closely associated with PSI in state 2 conditions, evoking the possible formation of supercomplexes in the latter condition. On the basis of analysis of the kinetics of cyt*f* oxidation induced by a single saturating flash, Delosme [49] reported that its oxidation rate is highly dependent upon physiological conditions. Reversible transitions were observed between a slow oxidation rate ($t_{½}$ ~500 µs) in oxygenic conditions to a fast oxidation rate ($t_{½}$ ~80 µs) under anaerobic conditions. On the basis of the results of Wollman and Bulté [25], Delosme [49] suggested that, in state 2 conditions, the fast oxidation of cyt*f* was due by the higher proximity between primary and secondary donors within supercomplexes. In agreement with this proposal, Iwai et al [29] reported that the formation of such supercomplexes would be promoted in state 2 conditions (reviewed by Minegawa in [50]). Here we observed that *Chlamydomonas* mutants devoid of the STT7 kinase activity that controls state transition did not display the ability to trap 2 PC within a functional supercomplex grouping 1 PSI and 1 cyt$b_6f$. However, the BF4-F34 mutant that lacks peripheral antenna proteins but still shows STT7 activity [51], produced such functional supercomplexes in anaerobic conditions. Therefore, another phosphoprotein, substrate of the STT7 kinase may play a role in promoting this functional reorganization. Buchert et al. [30] found interactions between several auxiliary proteins reported to enhance CEF, and cyt$b_6f$ complexes. Among these are ANR1 [30, 34, 52] and the phosphoprotein PETO [53]. Takahashi et al [34] showed that PETO improved CEF and better-preserved co-fractionation of PSI with cyt$b_6f$ and CEF effectors in sucrose gradients after membrane solubilization. PETO is a bitopic protein, with two very bulky domains in each side of its transmembrane helix. The possible contribution of PETO to trapping PC on its luminal side should be considered as a means to bridge PSI and cyt$b_6f$ in a supramolecular configuration that may not involve interactions between their transmembrane domains. Such arrangement may lead to the loss of the supramolecular structure upon detergent solubilization of the thylakoid membranes and explain the disparate results for the finding of supramolecular entity between PSI and cyt$b_6f$ by biochemical means. Thus, the hypothetical megastructures corresponding to the supercomplex functional configuration, would easily be dissociated by most of the current purification and solubilization procedure. In our view, only a structural analysis of membranes close to their native state would be able to assess the structural basis for the supercomplex functional configuration that we documented in this study.

**Declaration of competing interest**

The authors declare that they have no known competing financial interests or personal relationships that could have appeared to influence the work reported in this paper.

**Acknowledgements**

This work was supported by UMR7141, CNRS/ Sorbonne Université and by the Agence Nationale de la Recherche (Labex Dynamo : ANR-11-LABX-0011-01).